\documentclass[11pt, a4paper]{article}
\usepackage[textwidth = 7in, textheight=25cm,nohead]{geometry}
\usepackage{ulem}
\usepackage{graphicx, amssymb, amsmath, amsthm, amstext, booktabs, float, multirow, subfigure, rotating,natbib}
\usepackage{setspace}
\numberwithin{equation}{section} 
\numberwithin{figure}{section} 
\numberwithin{table}{section}

\bibpunct{(}{)}{,}{a}{,}{,}

\def\ppn{\vskip 6pt \noindent }
\def\R{{\mathbb{R}}}

\def\E{{\mathbb{E}}}
\newcommand{{\Xs}}{{\cal X}}
\newcommand{{\Ys}}{{\cal Y}}
\newcommand{{\Ls}}{{\cal L}}
\newcommand{{\Ss}}{{\cal S}}
\newcommand{{\Gs}}{{\cal G}}
\newcommand{{\Hs}}{{\cal H}}
\newcommand{{\Ns}}{{\cal N}}
\newcommand{{\Is}}{{\cal I}}
\newcommand{{\Bs}}{{\cal B}}
\newcommand{{\Cs}}{{\cal C}}
\newcommand{{\Rs}}{{\cal R}}
\newcommand{{\Us}}{{\cal U}}
\newcommand{{\pp}}{{\mathbf p}}
\newcommand{{\KK}}{{\mathbf K}}
\newcommand{{\HH}}{{\mathbf H}}
\newcommand{{\II}}{{\mathbf I}}
\newcommand{{\yy}}{{\mathbf y}}
\newcommand{{\ab}}{{\mathbf a}}

\newcommand{{\toL}}{{\overset{\mathcal{L}}{\longrightarrow}\ }}

\newcommand{{\dou}}{$\leadsto$\ }

\newcommand{{\phphu}}{{\phi(\Phi^{-1}(u))}}
\newcommand{{\phphv}}{{\phi(\Phi^{-1}(v))}}
\newcommand{{\phphUi}}{{\phi(\Phi^{-1}(\hat{U}_i))}}
\newcommand{{\phphVi}}{{\phi(\Phi^{-1}(\hat{V}_i))}}

\DeclareMathOperator{\var}{\mathbb{V}ar}

\begin{document}
\title{Local-likelihood transformation kernel density estimation for positive random variables }
\author{\sc{Gery Geenens}\thanks{Corresponding author: ggeenens@unsw.edu.au, School of Mathematics and Statistics, UNSW Australia, Sydney, tel +61 2 938 57032, fax +61 2 9385 7123 }\\School of Mathematics and Statistics,\\ UNSW Australia, Sydney,  \and \sc{Craig Wang} \\ Swiss Federal Institute of Technology (ETH), \\ Z\"urich, Switzerland }
\date{\today}
\maketitle
\thispagestyle{empty} 

\begin{abstract}

\noindent The kernel estimator is known not to be adequate for estimating the density of a positive random variable $X$. The main reason is the well-known boundary bias problems that it suffers from, but also its poor behaviour in the long right tail that such a density typically exhibits. A natural approach to this problem is to first estimate the density of the logarithm of $X$, and obtaining an estimate of the density of $X$ using standard results on functions of random variables (`back-transformation'). Although intuitive, the basic application of this idea yields very poor results, as was documented earlier in the literature. In this paper, the main reason for this underachievement is identified, and an easy fix is suggested. It is demonstrated that combining the transformation with local likelihood density estimation methods produces very good estimators of $\R^+$-supported densities, not only close to the boundary, but also in the right tail. The asymptotic properties of the proposed `local likelihood transformation kernel density estimators' are derived for a generic transformation, not only for the logarithm, which allows one to consider other transformations as well. One of them, called the `probex' transformation, is given more focus. Finally, the excellent behaviour of those estimators in practice is evidenced through a comprehensive simulation study and the analysis of several real data sets. A nice consequence of articulating the method around local-likelihood estimation is that the resulting density estimates are typically smooth and visually pleasant, without oversmoothing important features of the underlying density.

\end{abstract}

\section{Introduction}\label{sec:intro}

Kernel density estimation is arguably the most popular nonparametric density estimation method. Totally flexible, it makes no prior assumption on the functional shape of the density to estimate, and really `let the data speak for itself'. Given a sample $\Xs=\{X_k\}_{k=1}^n$ drawn from an unknown distribution $F_X$ admitting a density $f_X$, the estimator takes the form 
\begin{equation} \hat{f}_X(x) = \frac{1}{nh} \sum_{k=1}^n K\left(\frac{x-X_k}{h}\right), \label{eqn:kde} \end{equation}
where $K$ is a `kernel' function, usually a symmetric probability density of unit variance, such as $\phi$ the standard normal density, and $h$ is positive number called `bandwidth' which fixes the smoothness of the resulting estimate. The statistical properties of estimator (\ref{eqn:kde}) have been studied and understood for decades \citep{Wand95}, and the huge amount of theoretical and applied literature on kernel density estimation sufficiently testifies that it reliably estimates $f_X$ in a flexible way. Yet, this is actually true only if the support of $f_X$ is the whole real line $\R$. Boundaries in the support of $f$ generally cause much trouble for $\hat{f}_X$.

\ppn A case of bounded support of major importance is when $F_X$ is the distribution of a positive continuous random variable $X$, with density $f_X$ supported on $\R^+ = (0,+\infty)$. Those variables naturally arise in various areas of social, financial and medical sciences, such as banking transactions, travel durations, household incomes, or survival times, to cite only a few examples. Typically, those distributions are heavily skewed, with a density showing a maximum at or near the boundary 0 and a long tail on the right side. Often, the behaviour of the density close to 0 is what mostly matters for the analyst; in other cases, it is rather the tail behaviour of the distribution which is the main focus, for instance when high quantiles (e.g., Value-at-Risk) are of interest. Popular parametric models for this type of distributions include Exponential, Log-normal, Gamma, Weibull and Pareto. However, those parametric specifications are sometimes too rigid to appropriately model some random behaviour observed in practice. When the risk of misspecification is so high that positing such parametric models is hazardous, a totally flexible estimator such as (\ref{eqn:kde}) must be favoured. 

\ppn Yet, the kernel estimator (\ref{eqn:kde}) fails to correctly estimate both the behaviour of $f_X$ close to 0 and in the tail. Close to 0, the estimator suffers from {\it boundary bias}: not aware of the support boundary, $K((x-X_k)/h)$ typically overflows beyond it for the $X_k$'s close to 0, placing positive probability mass in forbidden areas. This results in an important bias which often prevents the estimator from being consistent there \citep[Section 2.11]{Wand95}. In the tail region, where data are usually sparse, it produces `{\it spurious bumps}' \citep{Hall04}, i.e.\ artificial local maxima at each observation, thus performing poorly as well.

\ppn In consequence, corrections, modifications and extensions of (\ref{eqn:kde}), attempting to make it suitable for $\R^+$-supported densities, abound in the literature. Early attempts at curing boundary effects looked for correcting $\hat{f}_X$ close to 0. These include, among others, the `cut-and-normalised' estimator \citep{Gasser79}, later refined in \cite{Jones93} and \cite{Jones96}, and the reflection method \citep{Schuster85}. These procedures are essentially {\it ad hoc} manual surgeries on (\ref{eqn:kde}) close to 0, and have, since then, shown their limitations both in theory and practice. In addition, they leave the tail area untouched, hence they do not address the `spurious bumps' phenomenon at all.

\ppn Later, the problem was rather addressed from a global perspective, by redefining the estimator as
\begin{equation} \hat{f}_X(x) = \frac{1}{n} \sum_{k=1}^n L(X_k;x,h), \label{eqn:asymkde} \end{equation}
where $L(\cdot;x,h)$ is an asymmetric $\R^+$-supported density whose parameters are functions of $x$ and $h$. By definition, such kernels $L$ cannot assign probability weight to the negative values, hence address the above boundary problems at the source. In addition, they typically share the same right-skewness as $f_X$ and this may be beneficial to the estimator in the tail area. Thus, asymmetric kernels appear to be the appropriate tool in this framework. A first attempt at this idea was made in \cite{Chen00} who took $L(\cdot;x,h)$ to be some Gamma density, defining two versions of the `Gamma kernel density estimator'. Although more types of asymmetric kernels, such as log-normal and Birnbaum-Saunders \citep{Jin03} or Inverse Gaussian and reciprocal Inverse Gaussian \citep{Scaillet04}, were investigated in the subsequent literature, those showed little advantage over Chen's Gamma kernel density estimator which arguably remains the main reference for asymmetric kernel estimation on $\R^+$. Its properties were further investigated in \cite{Bouezmarni05,Hagmann07,Zhang10} and \cite{Malec12}, and other ideas related to asymmetric kernel estimation were described in \cite{Kuruwita10,Comte12,Mnatsakanov12,Jeon13,Koul13,Marchant13,Igarashi14} and \cite{Igarashi16}. Recently, \cite{Hirukawa15} described a family of `generalised Gamma kernels' which includes a variety of similar asymmetric kernels in an attempt to standardise those results. 

\ppn This paper rather looks into a third type of procedure, seemingly different, based on data transformation as first suggested in \cite{Copas80} and \citet[Sections 2.9-2.10]{Silverman86}. The idea was explored in more details in \cite{Marron94} and \cite{Ruppert94}, and variants investigated in \cite{Koekemoer08} and \cite{Gustafsson09}, see also \cite{Wand91}. The idea is to transform, through some one-to-one continuous function $T$, the variable of interest $X$ into another one $Y=T(X)$ whose density estimation is not affected by boundary issues, and estimate the density of $X$ through that of $Y$. In the above papers, one has an `easy' distribution for $Y$ in mind, e.g.\ Uniform \citep{Marron94,Ruppert94,Gustafsson09} or Normal \citep{Koekemoer08}. Of course, the transformation $T$ able to produce that target density depends on the unknown $F_X$ and must be estimated from the data. In a sense, estimating $f_X$ is here substituted by estimating $T$. Unfortunately, estimating a transformation $T$ able to produce a given distribution is less natural and less convenient than estimating the distribution itself (or its density).

\ppn In this paper, the transformation idea is contemplated from a different perspective, totally in line with \cite{Copas80} and \cite{Silverman86}'s original suggestion. Consider a fixed, smooth and increasing function $T: \R^+ \to \R$ such that $\lim_{x \to 0} T(x) = -\infty$ and $\lim_{x \to +\infty} T(x)= +\infty$. Then, the distribution of $Y = T(X)$ is (under mild conditions) supported on the whole real line $\R$. Consequently, one should be able to estimate the density $f_Y$ of $Y$ free from boundary effects. Via back-transformation, one should obtain an estimate of $f_X$ enjoying that property, too. This differs from the ideas presented in the previous paragraph in the fact that here, no particular distribution for $Y$ is targeted. The transformation $T$ is fixed beforehand and, in a sense, just aims at sending the boundary away to $-\infty$. The so-produced variable $Y$ has some arbitrary distribution that must be estimated. Estimating $f_X$ is here substituted by estimating another density $f_Y$, which does not bring in any extra methodological or numerical burden. In fact, estimating $f_Y$ is usually easier, if only owing to the absence of boundary issues. 

\ppn \cite{Copas80}'s idea was to use the logarithmic transformation, which indeed seems natural for a positive random variable $X$. Define $Y = \log(X)$. From standard arguments, one has
\begin{equation*} 
 f_X(x) = \frac{f_Y(\log x)}{x} \label{eqn:fglog}
\end{equation*}
for all $x >0$, which readily suggests, upon estimation of $f_Y$ by an estimator $\hat{f}_Y$, an estimator of $f_X$:
\begin{equation} 
\hat{f}_X(x) = \frac{\hat{f}_Y(\log x)}{x}, \quad \text{ for } x >0. \label{eqn:fgloghat}
\end{equation}
Interestingly, if one uses a conventional kernel estimator (\ref{eqn:kde}) with Gaussian kernel ($K = \phi$) for estimating $f_Y$, i.e.
\begin{equation} \hat{f}_{Y}(y)=\frac{1}{nh}\sum_{k=1}^{n} \phi\left(\frac{y-Y_{k}}{h}\right) \label{eqn:fYhat1} \end{equation}
with $Y_k = \log X_k$, one obtains
\begin{equation} \hat{f}_X(x) = \frac{1}{nhx} \sum_{k=1}^n \phi\left(\frac{\log x-\log X_k}{h}\right) = \frac{1}{n} \sum_{k=1}^n \frac{1}{x\sqrt{2\pi h^2}}\exp\left(-\frac{(\log x-\log X_k)^2}{2h^2}\right). \label{eqn:lognormkde} \end{equation}
Clearly, this is equivalent to (\ref{eqn:asymkde}) with $L(\cdot;x,h)$ being the log-normal density with parameters $\mu = \log x$ and $\sigma = h$, an estimator which it seems fair to call the {\it log-normal kernel density estimator}.\footnote{Noticeably, it is different to \cite{Jin03}'s homonymous estimator. In fact, those authors use, for some unclear reason, the less natural parameterisation $\sigma = 2\log^{1/2}(1+h^2)$ for their log-normal kernel. Of course, $\log^{1/2}(1+h^2) \sim h$ as $h \to 0$, so the two parameterisations are asymptotically equivalent.} In fact, even for other choices than $T= \log$ and $K = \phi$, the two approaches (transformation and asymmetric kernels) can generally be regarded as two sides of the same idea, and are not that different after all.

\ppn In any case, \cite{Charpentier14} described the ability of (\ref{eqn:lognormkde}) to deal with skewed and heavy-tailed densities of positive random variables. They also admitted that it works well close to the boundary only if $f_X(0) = 0$, the reason for this being made clear in the next section. This restriction explains mostly why, although simple and intuitively appealing, this estimator has received little support in the literature: its practical performance is very disappointing in general, a good illustration of this being Figure 2.13 in \cite{Silverman86}. 

\ppn Recently, though, \cite{Geenens14} reconsidered the transformation idea in the related situation of a variable $X$ supported on $[0,1]$. In that paper, the main reasons for the previous failures of the transformation method were identified and some remedies were suggested. It turns out that estimating the density of the transformed variable $Y$ must be carried out with the greatest care. In particular, raw kernel estimation (\ref{eqn:fYhat1}) of $f_Y$ is not good enough to produce a good estimate of $f_X$ after back-transformation. In fact, the final estimate of $f_X$ in the boundary areas is extremely sensitive to any variability in the tails of the density estimate in the transformed domain. As a result, \cite{Geenens14} concluded that local likelihood methods, known to have superior tail behaviour when estimating a density, would be a better choice for estimating $f_Y$ with the final estimate of $f_X$ in mind. The suggested local likelihood-based transformation kernel density estimators were indeed seen to outperform their main competitors for densities supported on $[0,1]$. \cite{Geenens16} further demonstrated the potential of the method for estimating the density of a bivariate copula distribution on $[0,1]\times [0,1]$.

\ppn Here, the idea is investigated for the case where $f_X$ is supported on $\R^+$. As it was observed in the other situations, the suggested estimator will be seen to outperform all its competitors by a wide margin for a broad range of density shapes, close to the boundary, but also in the tail area. The paper is organised as follows. Section \ref{sec:transf} describes the transformation idea in more detail, explains why the `naive' approach based on (\ref{eqn:fYhat1}) does not work well, and introduces the transformation estimators based on local likelihood ideas. Section \ref{sec:asympt} derives the asymptotic properties of the suggested estimators. Section \ref{sec:logbey} discusses the choice of the transformation $T$, while Sections \ref{sec:NN} and \ref{sec:bw} focus on the always crucial problem of the smoothing parameter. Section \ref{sec:simul} consists of a comprehensive simulation study comparing the practical performance of the suggested estimators to that of their competitors. Section \ref{sec:realdata} briefly analyse some real data sets, and Section \ref{sec:ccl} concludes the paper and describes some research perspectives.

\section{Transformation Kernel Density Estimators} \label{sec:transf}

\subsection{Transformation} \label{subsec:trans}

The logarithm function seems the obvious choice for transforming a positive variable $X$ into another one $Y$ supported on the whole real line. However, other transformations can also do and are worth being considered, see Section \ref{sec:logbey}. Hence, the results below will be stated for an arbitrary transformation $T$, arguing that general results about such `Transformation Kernel Density Estimators' might be of interest on their own, too. So, let $T: \R^+ \to \R$ be a smooth (i.e., as many times differentiable as required) and increasing function, such that $\lim_{x \to 0^+} T(x) = -\infty$ and $\lim_{x \to +\infty} T(x)= +\infty$. Clearly, these conditions imply that $T'(x) = \frac{dT}{dx}(x)>0$ for all $x \in \R^+$, and that the inverse transformation $T^{-1}:\R \to \R^+$ is unequivocally defined. 

\ppn Define $Y= T(X)$ the random variable of interest in the transformed domain. Well-known results on functions of random variables state that, for all $y \in \R$,
\begin{equation} f_Y(y) = \frac{f_X(T^{-1}(y))}{T'(T^{-1}(y))}. \label{eqn:fY} \end{equation}
Obviously, if $f_X(x)>0$ a.e.\ on $\R^+$ - what will be assumed throughout the paper - $Y$ has unbounded support, and one should be able to estimate $f_Y$ free from any boundary issue. Inverting the above expression, one can also write that, for all $x > 0$,
\begin{equation} f_X(x) = f_Y(T(x)) \times T'(x). \label{eqn:fXT} \end{equation}

\ppn The original sample $\Xs$ can be readily transformed to obtain $\Ys=\{Y_k = T(X_k)\}^{n}_{k=1}$, a sample from $F_Y$, which can be used to estimate its density. Any estimator of $f_Y$, say $\hat{f}_Y$, automatically provides an estimator of $f_X$:
\begin{equation} \hat{f}^{(T)}_X(x) = \hat{f}_Y(T(x)) \times T'(x), \label{eqn:hatfXT} \end{equation}
where the superscript $(T)$ refers to the idea of transformation. The estimator can also be defined, if necessary, at $x=0$ by continuity: $\hat{f}^{(T)}_X(0) = \lim_{y \to -\infty} \hat{f}_Y(y) \times T'(T^{-1}(y))$. Clearly, $\hat{f}^{(T)}_X$ cannot allocate any probability mass to the negative side, as $T(x)$ is not defined for $x < 0$. Also, if the estimate $\hat{f}_Y$ is a {\it bona fide} density, in the sense that $\hat{f}_Y(y) \geq 0$ for all $y \in \R$ and $\int_\R \hat{f}_Y(y)\,dy = 1$, so automatically is $\hat{f}^{(T)}_X$. Finally, if $\hat{f}_{Y}$ is a uniformly (weak or strong) consistent estimator for $f_{Y}$, i.e.\ $\sup_{y \in \R} |\hat{f}_{Y}(y)-f_{Y}(y)| \overset{P/a.s.}{\to} 0$ as $n \to \infty$, $\hat{f}^{(T)}_X$ inherits that uniform consistency on any compact proper subset of $\R^+$. It seems, therefore, enough to find a good estimator for $f_Y$.

\ppn The tricky point is that a good estimator $\hat{f}_Y$ for estimating $f_Y$, is not necessarily the one that would produce a good estimator of $f_X$ through (\ref{eqn:hatfXT}). The main reason for this is that $\lim_{x \to 0} T(x) = -\infty$ together with $T$ smooth and increasing requires $\lim_{x \to 0} T'(x) = + \infty$, which is a cause of concern in (\ref{eqn:hatfXT}) when $x$ approaches 0. In particular, the estimator $\hat{f}_Y$ should mimic the behaviour of $f_Y$ in its left tail {\it very} accurately, given that any estimation error there will be greatly magnified by multiplying by a potentially huge $T'(x)$ when back in the initial domain. Hence, loosely speaking, a good estimator $\hat{f}_Y$ to be used in (\ref{eqn:hatfXT}) should be one which focuses more on its left tail than on its right tail. 

\subsection{Naive estimator} \label{subsec:naive}

\begin{figure}
\centering
\includegraphics[width=0.7\textwidth, trim=0 1.5cm 0 0]{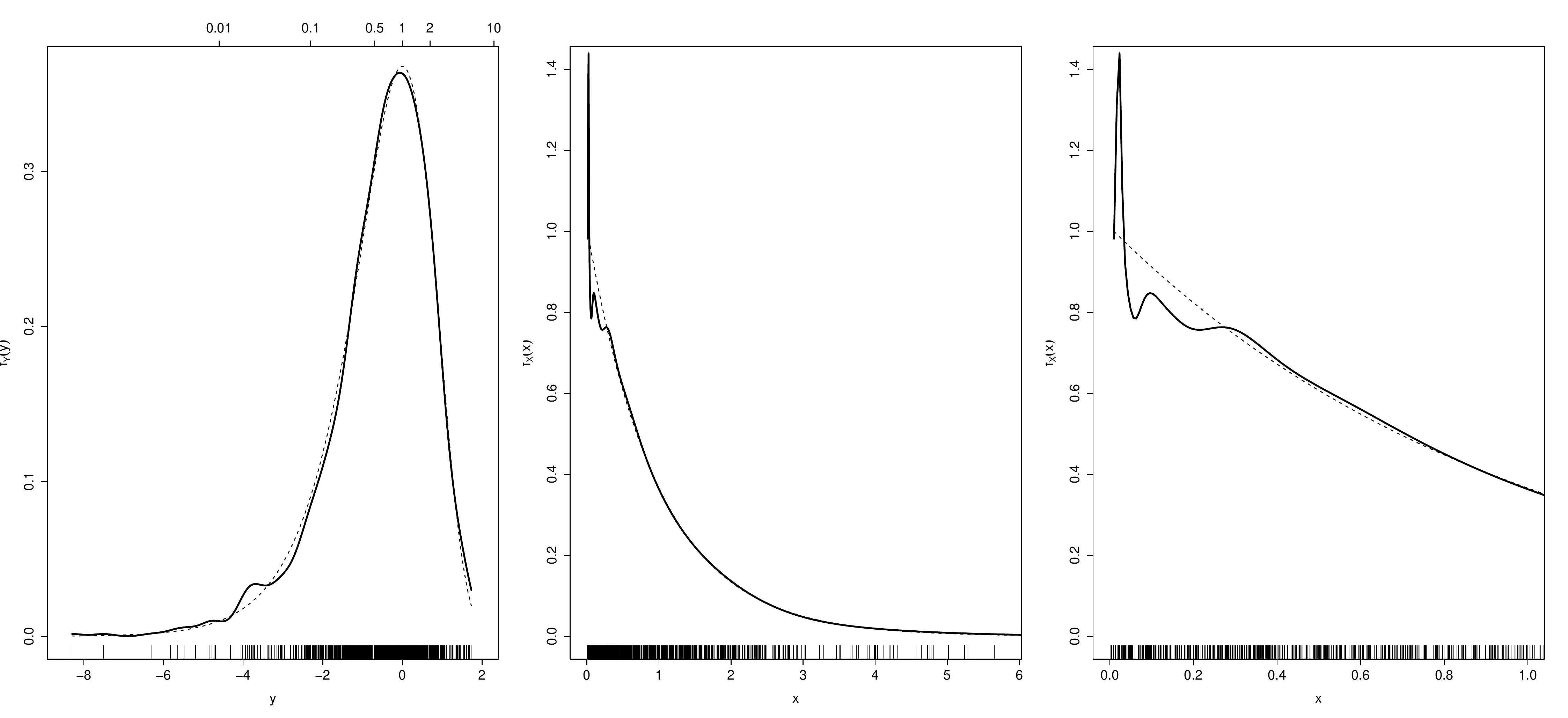}
\caption{The `naive' estimator $\hat{f}_{X}^{(T)}(x)$ (plain line, middle and right panels) computed on a simulated data set of size $n=1,000$ from the Exp$(1)$-density (dashed line). The left panel shows the data, the true and the estimated densities in the log-transformed domain.}
\label{fig:naiveex1}
\end{figure}

As in (\ref{eqn:fYhat1})-(\ref{eqn:lognormkde}), it seems - at first sight - natural to estimate $f_Y$ with the kernel estimator
\begin{equation} \hat{f}_{Y}(y)=\frac{1}{nh}\sum_{k=1}^{n}K\left(\frac{y-Y_{k}}{h}\right).\label{eqn:fYhat}\end{equation}
However, \cite{Geenens14} called this approach `naive', and the last paragraph of the previous subsection explains why. Indeed, conventional kernel estimators like (\ref{eqn:fYhat}) are known to poorly estimate tails of densities. Figure \ref{fig:naiveex1} illustrates the issues that this causes in the particular case of the log-normal kernel estimator (\ref{eqn:fgloghat})-(\ref{eqn:lognormkde}), but the same observations would hold for the general transformation estimator (\ref{eqn:hatfXT}) coupled with (\ref{eqn:fYhat}).

\ppn A random sample of size $n =1000$ from the Exponential(1) distribution was simulated (ticks in the middle panel), and transformed through the function $T(x) = \log x$ into a sample $\Ys$. By (\ref{eqn:fY}), $f_Y(y) = \exp(y - \exp(y))$ is here the Gumbel density on $\R$, which is estimated by (\ref{eqn:fYhat1}) with bandwidth $h=0.264$ selected by direct plug-in \citep{Sheather91}. This estimator is doing very well at reproducing the peak region of $f_Y$  (left panel), however in the left tail area `spurious bumps' show up. One is obvious around $y=-4$, but others are perceptible further in the tail as well. The corresponding estimate of $f_X$, obtained through back-transformation (\ref{eqn:fgloghat}), is shown in the middle panel. Away from the boundary, it is doing fine, but close to 0 it is extremely rough. The right panel shows its behaviour on the interval $[0,1]$. In fact, the `spurious bumps' in $\hat{f}_Y$ are greatly magnified by (\ref{eqn:fgloghat}) when $x$ is close to 0. In addition, the $\log$-function bluntly crushes that tail back to the positive side (the log-scaling on top of the left panel illustrates the extent of that crush), which explains why the bumps occur with higher and higher frequency when approaching 0, as seen in the right panel. Clearly, one cannot be happy with this estimate.

\subsection{Local-likelihood-based estimators} \label{subsec:LLest}

Particular care is thus required for estimating the left tail of $f_Y$, as this one will produce the estimate of $f_X$ near 0, where problems usually arise. Unlike raw kernel estimators, local likelihood (LL) density estimators \citep{Loader96,Hjort96,Park02} are known to have superior tail behaviour, producing much less wiggly estimates in the tail areas. Hence they seem totally appropriate in this framework. \cite{Loader96}'s local likelihood estimator is defined as follows. Around $y \in \R$, it is assumed that the log-density can be well approximated by some polynomial of degree $p$, i.e., for $\check{y}$ `close' to $y$,
\[\log f_Y(\check{y}) \simeq a_0(y) + a_1(y) (\check{y}-y) + \ldots + a_p(y) (\check{y}-y)^p. \]
Then, making use of this approximation, a local maximum likelihood problem is solved at $y$:
\begin{multline} (\tilde{a}_0(y),\ldots,\tilde{a}_p(y)) = \arg \max_{a_0,a_1,\ldots,a_p}\left\{ \sum_{i=1}^n K\left(\frac{Y_i-y}{h}\right) \left(a_0 + a_1 (Y_i-y) +\ldots + a_p (Y_i-y)^p \right)\right.\\ \left.- n \int_\R K\left(\frac{t-y}{h}\right)\exp\left(a_0 + a_1 (t-y) +\ldots + a_p (t-y)^p\right)\,dt\right\}. \label{eqn:loclikpol} \end{multline}
The estimate of $f_Y$ at $y$ is finally obtained as $\tilde{f}_Y^{(p)}(y) = \tilde{a}_0(y)$. Typically, only the cases $p=1$ (local log-linear) and $p=2$ (local log-quadratic) are considered. Essentially, local log-linear estimation forces $\tilde{f}_Y^{(1)}$ to behave {\it locally } like a certain Exponential density, whereas the log-quadratic version drives $\tilde{f}_Y^{(2)}$ so as to behave {\it locally} as a certain Normal density (`local parametric modelling', \citet{Hjort96}). As a result, these estimators generally produce very smooth estimates, in particular in the tails. Using these $\tilde{f}_Y^{(p)}$ instead of (\ref{eqn:fYhat}) in (\ref{eqn:hatfXT}) should, therefore, correct the defects described in the previous subsection. The final, local-likelihood transformation kernel density estimators (LLTKDE) will be given at any $x > 0$ by 
\begin{equation} \tilde{f}_X^{(T,p)}(x) = \tilde{f}_Y^{(p)}(T(x)) \times T'(x) \qquad (p=1,2). \label{eqn:ftildeTp} \end{equation}
The maximisation problem (\ref{eqn:loclikpol}) is easily realised by the R functions available in the {\tt locfit} package. In the next section, the asymptotic properties of these transformation estimators (`naive' and `LL') are studied for a generic transformation $T$. 
  
\section{Asymptotic properties} \label{sec:asympt}

Below, the kernel $K$ in (\ref{eqn:fYhat}) as well as in (\ref{eqn:loclikpol}) will be assumed to be a symmetric probability density function with $\int u^j K(u)\,du = \mu_j$ and $\int u^j K^2(u)\,du = \nu_j$ for $j=0,1,2\ldots$. The bandwidth $h$ will be such that $h \to 0$ and $nh \to \infty$ as $n \to \infty$. This is the classical framework in kernel density estimation.

\subsection{Naive estimator} 

Suppose that $f_Y$ is two times continuously differentiable at $y \in \R$. Then, it is well known that the kernel estimator (\ref{eqn:fYhat}) is such that
\begin{equation} \sqrt{nh}\left(\hat{f}_Y(y) - f_Y(y) -\frac{1}{2}\mu_2h^2 f''_Y(y)\right) \toL \Ns\left(0,\nu_0 f_Y(y)\right), \label{eqm:asymptYnaive}\end{equation}
for $h = O(n^{-1/5})$ as $n \to \infty$. Hence, 
\[\sqrt{nh}T'(x)\left(\hat{f}_Y(T(x)) - f_Y(T(x)) -\frac{1}{2}\mu_2h^2 f''_Y(T(x))\right) \toL \Ns\left(0,T'^2(x)\nu_0 f_Y(T(x))\right), \]
for any $x>0$ such that $f_Y$ is twice continuously differentiable at $T(x)$. Through (\ref{eqn:fXT})-(\ref{eqn:hatfXT}), this directly implies that 
\begin{equation} \sqrt{nh}\left(\hat{f}^{(T)}_X(x) - f_X(x) -\frac{1}{2}\mu_2h^2 b_T(x)\right) \toL \Ns\left(0,\nu_0 v_T^2(x)\right), \label{eqn:asymptnaive} \end{equation}
where $b_T(x) = T'(x)f''_Y(T(x))$ and $v_T^2(x) = T'^2(x)f_Y(T(x))$, i.e.\ 
\begin{equation}  v_T^2(x) =  T'(x) f_X(x), \label{eqn:varT} \end{equation}
from (\ref{eqn:fXT}). As $\lim_{x \to 0} T'(x) = \infty$, the variance of $\hat{f}^{(T)}_X$ is seen to grow unboundedly towards the boundary, unless $f_X$ tends to 0 {\it very smoothly} as $x \to 0$ (in the sense $f_X(x) = O(1/T'(x)$). This explains the extremely wiggly behaviour of the estimate near 0 in Figure \ref{fig:naiveex1}. Now, differentiating (\ref{eqn:fY}) one gets
\begin{equation}
 f'_Y(y)  = \frac{f_X'(T^{-1}(y)}{T'^2(T^{-1}(y))} - \frac{f_X(T^{-1}(y))T''(T^{-1}(y))}{T'^3(T^{-1}(y))} \label{eqn:fYprime} \end{equation}
 \begin{multline}
 \text{and} \quad  f''_Y(y)  = \frac{f''_X(T^{-1}(y))}{T'^3(T^{-1}(y))} - 3\, \frac{f'_X(T^{-1}(y))T''(T^{-1}(y))}{T'^4(T^{-1}(y))} \\  - \frac{f_X(T^{-1}(y))T'''(T^{-1}(y))}{T'^4(T^{-1}(y))} + 3\,\frac{f_X(T^{-1}(y))T''^2(T^{-1}(y))}{T'^5(T^{-1}(y))}. \label{eqn:fYsecond}
 \end{multline} 
Hence, the asymptotic bias of $\hat{f}^{(T)}_X(x)$ is $\frac{1}{2} \mu_2 h^2 b_T(x)$ where 
\begin{equation} b_T(x) = \frac{f''_X(x)}{T'^2(x)} - 3\, \frac{f'_X(x)T''(x)}{T'^3(x)}+f_X(x) \left(3\,\frac{T''^2(x)}{T'^4(x)} -\frac{T'''(x)}{T'^3(x)}  \right). \label{eqn:biasnaive} \end{equation}
Unlike what one usually has, the bias of the `naive' transformation kernel density estimator $\hat{f}_X(x)$ does not only involve $f''_X(x)$, but also $f'_X(x)$ and $f_X(x)$ itself, which is obviously not desirable.

\ppn For the particular case $T(x) = \log x$, $1/T'^2(x) = x^2$, $-3T''(x)/T'^3(x) = 3x$ and $\left(3\,\frac{T''^2(x)}{T'^4(x)} -\frac{T'''(x)}{T'^3(x)}  \right) \equiv 1$, so that
\begin{equation*} b_{\log}(x) = x^2 f''_X(x) + 3x f'_X(x) + f_X(x). \label{eqn:biasnaivelog} \end{equation*}
In the same time,
\[v_{\log}^2(x) =  \frac{f_X(x)}{x}. \] 
As soon as $f_X(0) >0$, this `naive' log-transform estimator, i.e.\ the log-normal kernel estimator (\ref{eqn:lognormkde}), will thus show both substantial bias and variance at the boundary, which explains \cite{Charpentier14}'s comments and the little support it has got. The general expression (\ref{eqn:biasnaive}) leaves little hope to find another transformation $T$ that would do much better. For instance, the general solution of the o.d.e.\ $3\,\frac{T''^2(x)}{T'^4(x)} -\frac{T'''(x)}{T'^3(x)} \equiv 0$, that would automatically cancel the last term in $b_T(x)$, is $T(x) = \sqrt{C_1 x + C_2}$, for any two constants $C_1$ and $C_2$. But this transformation is not such that $\lim_{x \to 0^+} T(x) = -\infty$ as prescribed, hence is not useful for solving the boundary issues.

\subsection{Local-likelihood-based estimators: log-linear case} \label{subsec:loclin}

The asymptotic properties of the local-likelihood-based transformation estimators can be derived in a similar way. From \cite{Loader96}'s results, the local log-linear estimator ($p=1$ in (\ref{eqn:loclikpol})) is such that
\begin{equation} \sqrt{nh} \left( \tilde{f}_Y^{(1)}(y) -f_Y(y) - \frac{1}{2}\mu_2 h^2 \left(f''_Y(y) - \frac{f'^2_Y(y)}{f_Y(y)} \right) \right) \toL \Ns\left(0, \nu_0 f_Y(y) \right) \label{eqm:asymptYloclin} \end{equation} 
at any $y \in \R$ at which $f_Y$ is positive and twice continuously differentiable. Note that $f_Y(y) >0$ is guaranteed by (\ref{eqn:fY}) as soon as $f_X(T^{-1}(y)) >0$, what is assumed to be the case ($f_X(x) >0$ a.e.\ on $\R^+$). 
With (\ref{eqn:fYprime}) and (\ref{eqn:fYsecond}), it easily follows that the local log-linear transformation estimator $\tilde{f}^{(T,1)}_X$ (\ref{eqn:ftildeTp}) is such that
\begin{equation} \sqrt{nh} \left( \tilde{f}_X^{(T,1)}(x) -f_X(x) - \frac{1}{2}\mu_2 h^2 b_T^{(1)}(x) \right) \toL \Ns\left(0, \nu_0 v_T^{2}(x) \right) \label{eqn:asymptloclin} \end{equation}
at any $x>0$ at which $f_X$ is positive and twice continuously differentiable, with 
\begin{equation} b_T^{(1)}(x) =  \frac{1}{T'^2(x)}\,\left(f''_X(x) - \frac{f'^2_X(x)}{f_X(x)} \right) - \frac{f'_X(x)T''(x)}{T'^3(x)}+f_X(x) \left(2\,\frac{T''^2(x)}{T'^4(x)} -\frac{T'''(x)}{T'^3(x)}  \right) \label{eqn:biastilde1} \end{equation}
and $v^{2}_T(x) = T'(x) f_X(x)$. This asymptotic variance is the same as that of the `naive' version. In particular, it still grows unboundedly for $x$ approaching 0 when $f_X$ does not tend fast enough to 0 there. The bias component, on the other hand, is slightly different. 

\ppn For this local log-linear estimator, the choice $T(x) = \log x$ seems more justified than for the naive version. Indeed, $2\,T''^2(x)-T'(x)T'''(x)  \equiv 0$ for that choice, yielding 
\[b_{\log}^{(1)}(x) =  x^2\left(f''_X(x) - \frac{f'^2_X(x)}{f_X(x)} \right) - xf'_X(x). \]
The $\log$-transform thus automatically deals with the term proportional to $f_X$ in (\ref{eqn:biastilde1}). In addition, provided that $f''_X(x) - \frac{f'^2_X(x)}{f_X(x)}$ and $f'_X(x)$ remain bounded as $x\to 0$, the bias will actually be of order $o(h^2)$ at the boundary. If $f_X$ is bounded, then the bias is always at most $O(h^2)$, including at the boundary. In addition, if $f_X(x)/x =O(1)$ as $x \to 0$, then the variance is $O((nh)^{-1})$ everywhere as well. Balancing squared bias and variance, one can see that the local log-linear transformation kernel estimator attains the optimal rate of convergence, i.e.\ $O(n^{-4/5})$, for twice continuously differentiable densities. The exact rate of convergence for the other cases ($f_X(x)/x \to \infty$ as $x \to \infty$) would depend on the exact behaviour of $f_X$ close to the boundary.

\subsection{Local-likelihood-based estimators: log-quadratic case} \label{subsec:locquadra}

Deriving the properties of $\tilde{f}^{(T,2)}_X$ is in many points similar to the above, except that the bias expressions soon become rather unwieldy. In fact, it is well known in local polynomial modelling that fitting to a higher degree, here taking $p=2$ in (\ref{eqn:loclikpol}), usually reduces the order of the bias from order $O(h^2)$ to $O(h^4)$, smoothness of the underlying density permitting \citep{Fan96,Hjort96,Loader96}. Specifically, \cite{Loader96}'s and \cite{Hjort96}'s results show that, for $y \in \R$ at which $f_Y$ is positive and 4 times continuously differentiable, 
\begin{equation} \sqrt{nh} \left(\tilde{f}_Y^{(2)}(y) - f_Y(y) - \frac{1}{24}\frac{\mu_2 \mu_6 - \mu_4^2}{\mu_4 - \mu_2^2}h^4b^{(2)}_Y(y) \right)  \toL \Ns\left(0, V_2 f_Y(y) \right), \label{eqm:asymptYquadra} \end{equation} 
where $V_2 = \frac{\mu_4^2 \nu_0 -2\mu_2 \mu_4 \nu_2 + \mu_2^2 \nu_4}{(\mu_4 - \mu_2^2)^2}$ and 
\[b^{(2)}_Y(y) = f''''_Y(y) - 3\,\frac{f''^2_Y(y)}{f_Y(y)}+2\,\frac{f'^4_Y(y)}{f^3_Y(y)}.\]
Now, (\ref{eqn:fYsecond}) can be differentiated further to obtain the first four derivatives of $f_Y$ in terms of the first four derivatives of $f_X$ and the first five derivatives of $T$. Tedious algebraic developments eventually yield that, for any $x > 0$ at which $f_X$ is positive and four times continuously differentiable, 
\begin{equation} \sqrt{nh} \left(\tilde{f}_X^{(T,2)}(x) - f_X(x) - \frac{1}{24}\frac{\mu_2 \mu_6 - \mu_4^2}{\mu_4 - \mu_2^2}h^4b_T^{(2)}(x) \right)  \toL \Ns\left(0, V_2 v_T^{2}(x) \right), \label{eqn:asymptquadra} \end{equation}
where $v_T^2(x) = T'(x) f_X(x)$ and 
\begin{align}  b_T^{(2)}(x) =\  & \frac{1}{T'^4(x)}\left(f''''_X(x) - 3\,\frac{f''^2_X(x)}{f_X(x)} + 2\,\frac{f'^4_X(x)}{f^3_X(x)} \right) \notag \\ 
& - \frac{2T''(x)}{T'^5(x)}\left(5 f'''_X(x) -\frac{  9f'_X(x)f''_X(x)}{f_X(x)} + \frac{4f'^3_X(x)}{f_X^2(x)}\right)  + \frac{3T''^2(x)}{T'^6(x)}\left(9 f''_X(x) -\frac{ 5f'^2_X(x)}{f_X(x)}\right)  \notag \\ &- 4\,\frac{T'''(x)}{T'^5(x)}f''_X(x)- \left(59\,\frac{T''^3(x)}{T'^7(x)}-42\,\frac{T''(x)T'''(x)}{T'^6(x)}+5\,\frac{T''''(x)}{T'^5(x)} \right) f'_X(x) \notag \\
& + \left(80\,\frac{T''^4(x)}{T'^8(x)} - 87\,\frac{T''^2(x)T'''(x)}{T'^7(x)} + 7\,\frac{T'''^2(x)}{T'^6(x)}+15\,\frac{T''(x)T''''(x)}{T'^6(x)}-\frac{T'''''(x)}{T'^5(x)} \right) f_X(x). \label{eqn:bT2}
\end{align}
Compared to the naive or the log-linear estimator, the variance is multiplied by $V_2 >1$. E.g., if $K = \phi$ the Gaussian kernel, then $V_2 = 27/16$. This `inflation' factor is a well-known feature in local polynomial modelling when fitting a higher order polynomial \citep[Section 3.3.1]{Fan96}, and is the price to pay for reducing the order of the bias.

\ppn The bias expression (\ref{eqn:bT2}) seems hardly interpretable. Yet, taking $T(x) = \log x$ automatically cancels the last term again. For this choice, the above expression reduces to 
\begin{align*}  b_{\log}^{(2)}(x) =\  & x^4\left(f''''_X(x) - 3\,\frac{f''^2_X(x)}{f_X(x)} + 2\,\frac{f'^4_X(x)}{f^3_X(x)} \right)  +2x^3\left(5 f'''_X(x) -\frac{  9f'_X(x)f''_X(x)}{f_X(x)}  + \frac{4f'^3_X(x)}{f_X^2(x)}\right) \\ & + x^2\left(19 f''_X(x) -\frac{ 15f'^2_X(x)}{f_X(x)}\right)  +5x f'_X(x).
\end{align*}
Now, provided that each of the expressions between brackets remain bounded when approaching 0, the bias is actually $o(h^4)$ at the boundary. In any case, if $f_X$ is bounded, then the bias is $O(h^4)$, everywhere. As for the local log-linear case, the variance is $O((nh)^{-1})$ including at the boundary if $f_X(x)/x =O(1)$ as $x \to 0$. Hence, under this condition, the rate of convergence of the local log-quadratic transformation kernel estimator would be, this time, $O(n^{-8/9})$ uniformly on $\R^+$. This rate is the optimal convergence rate for nonparametrically estimating 4 times continuously differentiable densities. Of course, the estimator would always achieve this rate away from the boundary, regardless of the behaviour of $f_X$ as $x \to 0$.

\subsection{Dependent observations} \label{subsec:dependent}

Note that the asymptotic normality statements (\ref{eqn:asymptnaive})-(\ref{eqn:asymptloclin})-(\ref{eqn:asymptquadra}) not only hold for an i.i.d.\ sample $\Xs=\{X_k\}_{k=1}^n$. Those results directly follow from the corresponding expressions in the transformed domain ((\ref{eqm:asymptYnaive}), (\ref{eqm:asymptYloclin}) and (\ref{eqm:asymptYquadra}), respectively), where the usual kernel and local likelihood estimators are used. Yet, it happens that (\ref{eqm:asymptYnaive}), (\ref{eqm:asymptYloclin}) and (\ref{eqm:asymptYquadra}) hold true in the presence of dependence in the sample as well. This is well-known for the conventional kernel density estimator \citep{Robinson83}, and \cite{Tenreiro99} showed that this is also the case for local-likelihood density estimators if $\Ys = \{Y_k\}_{k=1}^n$ is a stationary $\alpha$-mixing (i.e.\ strongly-mixing) sequence with mixing coefficient $\alpha(m) = O(m^{-\omega})$, $\omega >2$. See also  and \cite{Lee04} for similar results. From \citet[Lemma 2.1]{White84} or \citet[Lemma 2.2.1]{Fan90}, $\Ys$ is such an $\alpha$-mixing sequence if $\Xs=\{X_k\}_{k=1}^n$ is so. Given that many stochastic processes and time series observed in practice are known to be $\alpha$-mixing \citep{Doukhan94}, the estimators proposed in this paper can be used for estimating marginal densities of most positive time series as well, with unchanged asymptotic properties.

\section{Beyond the log-transformation} \label{sec:logbey}

Although natural, the log-transformation need not be the best choice for $T$ in this framework. First, it is seen that the variance of the three estimators $\hat{f}^{(T)}_X$, $\tilde{f}_X^{(T,1)}$ and $\tilde{f}_X^{(T,2)}$ is proportional to $T'(x)$. So, it might be interesting to use a function $T$ which plunges to $-\infty$ as slowly as possible as $x \to 0$, in order to lower the variance of the estimator in the boundary area. Second, it can be understood that the local log-quadratic estimator $\tilde{f}_Y^{(2)}$ is particularly good at recovering a normal density in the transformed domain. This is because the `local parametric model' is right, in the words of \cite{Hjort96}: $\log \phi_{\mu,\sigma}(x)$ is indeed a quadratic function, for $\phi_{\mu,\sigma}$ being the $\Ns(\mu,\sigma^2)$-density. \cite{Hjort96} showed that the bias of the local log-quadratic estimator is actually $o(h^4)$ for normal densities. Now, the normal distribution is the maximum entropy distribution on $\R$ (for fixed mean and variance). In some sense, maximum entropy distributions are the `natural' distributions on a given domain, and it seems fair to expect a density estimator to be able to comfortably estimate them. The maximum entropy distribution on $\R^+$ is the exponential distribution. Through the probability integral transform, the transformation that would make the standard exponential distribution on $\R^+$ into the standard normal distribution on $\R$ is
\begin{equation} T(x) = \Phi^{-1}(1-e^{-x}), \label{eqn:probex} \end{equation}
where $\Phi^{-1}$ is the probit transformation, i.e.\ the inverse of the standard normal cumulative distribution function $\Phi$. Thus, using this transformation $Y = \Phi^{-1}(1-e^{-X})$ together with local log-quadratic estimation of $f_Y$, would make as easy to estimate an Exponential density on $\R^+$ as it is to estimate a Normal density on $\R$, which motivates its use. 

\begin{figure}[h]
\centering
\includegraphics[width=0.4\textwidth]{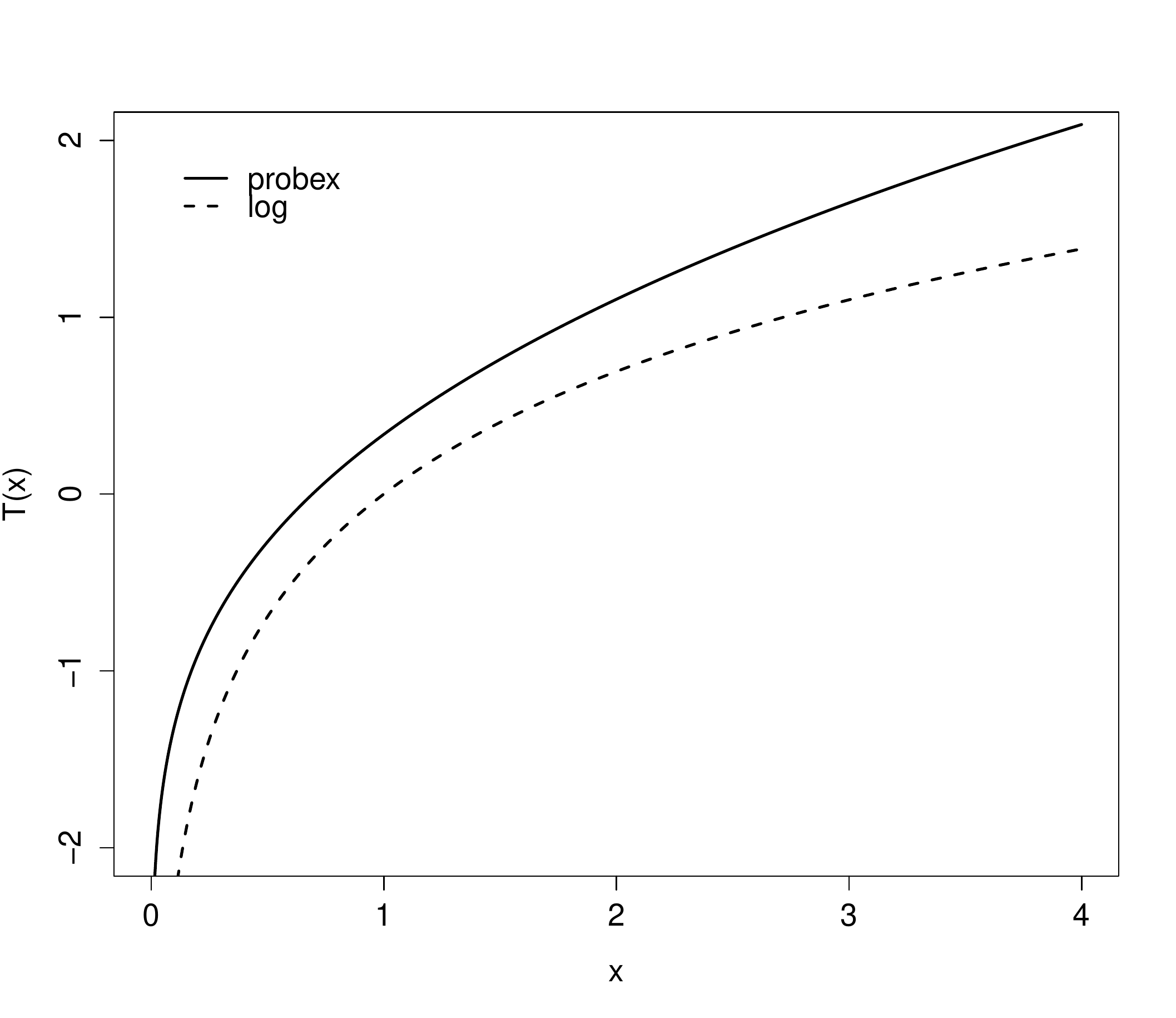}
\caption{Probex transformation function and log function.}
\label{fig:probex}
\end{figure}

\ppn This function (\ref{eqn:probex}), that could be called the {\it probex} transformation, is shown in Figure \ref{fig:probex}, together with the log function. The two transformations basically share the same appearance, but it can easily be checked that,
\[\frac{T'(x)}{(\log x)'} = \frac{x e^{-x}}{\phi(\Phi^{-1}(1-e^{-x}))} \to 0 \qquad \text{ as } x \to 0,\]
and $\frac{T'(x)}{(\log x)'}<1$ for $x \in (0,1)$. Thus, compared to the log-transform, the `probex' transformation indeed tends to stabilise the variance of the estimator close to the boundary. The first five derivatives of (\ref{eqn:probex}) could also be plugged in the bias expression (\ref{eqn:bT2}), but that would probably have little practical value. A practical comparison of the two transformations will rather be made in Section \ref{sec:simul}, through simulations. 

\ppn Note that this section is not an attempt to identify what the optimal transformation would be, or to argue in favour of choosing the transformation from the data. The idea of introducing the `probex' transformation is not to try to produce a variable $Y$ which is normally distributed: that would indeed be against the initial motivation of this paper, see Section \ref{sec:intro}. Rather, it is an invitation to considering other transformations than the logarithm. It is clear that many other transformations $T$ could be considered as well. Here, the idea was mainly to propose an estimator which can estimate Exponential-like densities with high accuracy. Indeed, many of the previous estimators are known to struggle when $f_X(0) >0$, of course including the `easy' Exponential case. 

\ppn For the reasons exposed above, the log-transform, used in conjunction with local log-quadratic estimation of $f_Y$, would produce an estimator particularly accurate for log-normal densities. This is attractive, too, however log-normal densities are always such that $f_X(0)=0$, so there is probably less need to focus on that case: most of the other estimators are able to deal with them reasonably well. In any case, the results in Sections \ref{sec:simul} and \ref{sec:realdata} show that the `probex'-transformation-based LLTKD estimator does also very well for densities such that $f_X(0)=0$.

\section{Nearest-Neighbour bandwidth} \label{sec:NN}

The previous developments have focused on the case of a fixed bandwidth $h$ in (\ref{eqn:loclikpol}), which is the classical choice. However, as observed earlier, the (asymptotic) variance of the LL-based transformation estimators is 
\[\var\left(\tilde{f}_X^{(T,p)}(x)\right) \sim V_p \frac{f_X(x)T'(x) }{nh},\]
with $V_1 = \nu_0$ and $V_2 = \frac{\mu_4^2 \nu_0 -2\mu_2 \mu_4 \nu_2 + \mu_2^2 \nu_4}{(\mu_4 - \mu_2^2)^2}$, as per (\ref{eqn:asymptloclin}) and (\ref{eqn:asymptquadra}). As $T'(x) \to \infty$ as $x \to 0$, it usually grows unboundedly when approaching the boundary, except in favourable situations. Even if using one or another transformation $T$ can slightly temper it (see Section \ref{sec:logbey}), this remains a major problem. In particular, this prevents the estimators from reaching optimal rates of convergence at the boundary in most of the cases, see discussion in Subsections \ref{subsec:loclin} and \ref{subsec:locquadra}.

\ppn In response to this, a simple solution seems to work with a Nearest-Neighbour (NN) type of bandwidth in the transformed domain. This is a {\it local} bandwidth, meaning that, in (\ref{eqn:loclikpol}), $h \doteq h(y)$ varies according to the value $y$ at which $f_Y$ is estimated. Specifically, $h(y)$ is set to $D_\alpha(y)=|y-Y_{(\alpha),y}|$, where $Y_{(\alpha),y}$ is the $\lfloor \alpha \times n \rfloor$th closest observation to $y$ out of the sample $\{Y_1,\ldots,Y_n\}$. It is now $\alpha$, the fraction of observations actively entering the estimation of $f_Y(y)$ at any $y$, which acts as smoothing parameter. Interestingly, as $T$ is a monotonic transformation, $\alpha$ is actually also the fraction of observations from the initial sample $\{X_1,\ldots,X_n\}$ actively entering the final estimation of $f_X(x)$ at any $x \in \R^+$.

\ppn The reason why this is advantageous in this setting, is the following. According to the above definition, $D_\alpha(y)$ is the distance between $y$ and its $\lfloor \alpha \times n \rfloor$th closest observation, hence it is random. The behaviour of such a quantity is well-understood, see \cite{Mack79} or \cite{Evans02}. In particular, it is known that, for $\alpha \to 0$, $n\alpha \to \infty$ as $n\to \infty$, 
\[\E(1/D_\alpha(y)) \sim \frac{2f_Y(y)}{\alpha}, \]
asymptotically. Using $\var(\tilde{f}^{(p)}_Y(y)) = \E\left(\var(\tilde{f}^{(p)}_Y(y))|D_\alpha(y) \right) + \var\left(\E (\tilde{f}^{(p)}_Y(y))|D_\alpha(y) \right)$, and the fact that, conditionally on $D_\alpha(y)$, the results of Subsections \ref{subsec:loclin}-\ref{subsec:locquadra} apply, straightforward manipulations lead to 
\[\var\left(\tilde{f}_X^{(T,p)}(x)\right) \sim 2V_p \frac{f^2_X(x)}{n\alpha}. \]
This expression is now free from any structurally unbounded factor. For bounded densities, the variance is thus, always, of order $O((n\alpha)^{-1})$. The estimators are obtained exactly as in Subsection \ref{subsec:LLest}, starting from (\ref{eqn:loclikpol}) with $h$ replaced by $D_\alpha(y)$. Not only this allows the variance of the estimators to be kept under control in the boundary region, it also provides the final LLTKD estimators with an adaptive nature all over $\R^+$. This way of doing has really proved efficient in \cite{Geenens14} and \cite{Geenens16}, and this will be confirmed in Sections \ref{sec:simul} and \ref{sec:realdata} in this setting as well. 

\ppn Note that this `trick' of using an NN-bandwidth cannot really be used for stabilising the variance of the `naive' estimator. In fact, a Nearest-Neighbour bandwidth works very well in conjunction with Local Likelihood estimation, but not so with traditional kernel methods. In the latter case, the produced estimates are typically very rough and show too fat tails, see Figure 2.10 in \cite{Silverman86}. In the former case, though, the estimates are forced to remain smooth, by the local parametric assumption, see discussion in \citet[Section 3.4]{Simonoff96}.

\section{Bandwidth selection} \label{sec:bw}

As is always the case in nonparametric modelling, the practical implementation of the above-described estimators requires the careful selection of a smoothing parameter, here $h$ or $\alpha$ in (\ref{eqn:loclikpol}). Classically, the main two approaches for selecting the bandwidth in kernel density estimation are {\it direct plug-in} \citep{Sheather91} and {\it cross-validation} \citep{Rudemo82,Bowman85}. Plug-in methods aim at estimating the (integrated) bias and variance of the estimator from their asymptotic expressions, and then picking the value of $h$ that minimise the resulting estimated integrated mean squared error. Given the appearance of the bias expressions (\ref{eqn:biastilde1}) and (\ref{eqn:bT2}), this approach seems out of reach in this situation, hence cross-validation is favoured. The smoothing parameter being selected in the transformed domain, the value of $h$ or $\alpha$ will be selected as the minimiser of the {\it least-square cross-validation} (LSCV) criterion
\begin{equation} \int_{-\infty}^{\infty} \left\{\tilde{f}_Y^{(p)}(y)\right\}^2 \,dy - \frac{2}{n} \sum_{k=1}^n \tilde{f}_{Y(-k)}^{(p)}(Y_k), \label{eqn:LSCV} \end{equation}
for $p =1$ or 2. As usual in a cross-validation situation, $\tilde{f}_{Y(-k)}^{(p)}$ is the `leave-one-out' version of the estimator, computed on all observations but $Y_k$. Expression (\ref{eqn:LSCV}) is known to be an unbiased estimate of the integrated squared error $\int_{-\infty}^{\infty} \left(\tilde{f}^{(p)}_Y(y) - f_Y(y)\right)^2\,dy$. The selected bandwidth should, therefore, be optimal for the data set at hand. \cite{Loader99} argues in favour of cross-validation methods over plug-in approaches, in particular when local likelihood methods are involved. This choice, therefore, seems appropriate here, and will exhibit very good performance in the simulation study in the next section. The LSCV function (\ref{eqn:LSCV}) is implemented in the {\tt locfit} package in R, allowing for both a fixed bandwidth $h$ and an NN-bandwidth $\alpha$.

\section{Simulation study} \label{sec:simul}

\begin{figure}[h]
\centering
\includegraphics[scale=0.5]{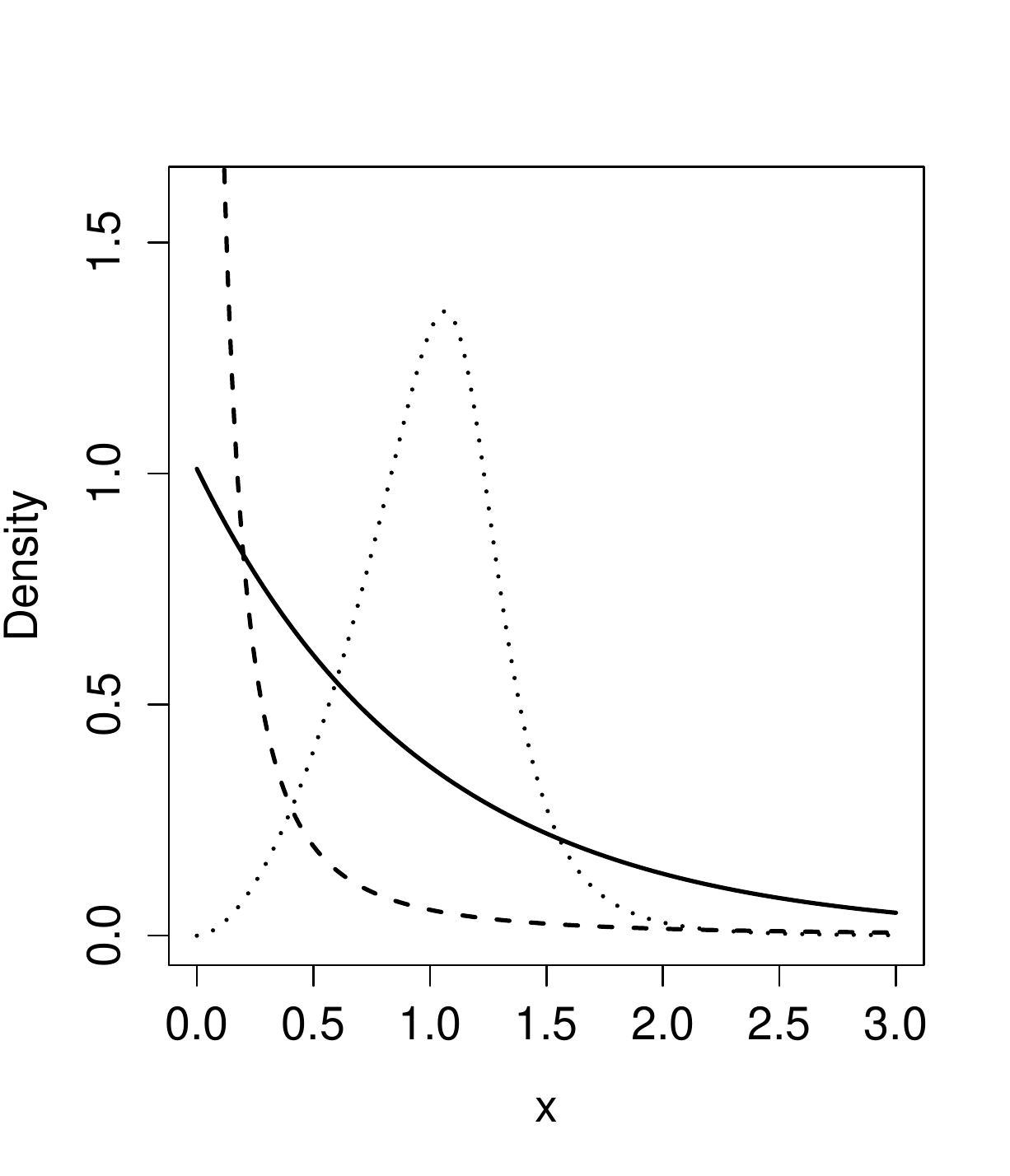}\includegraphics[scale=0.5]{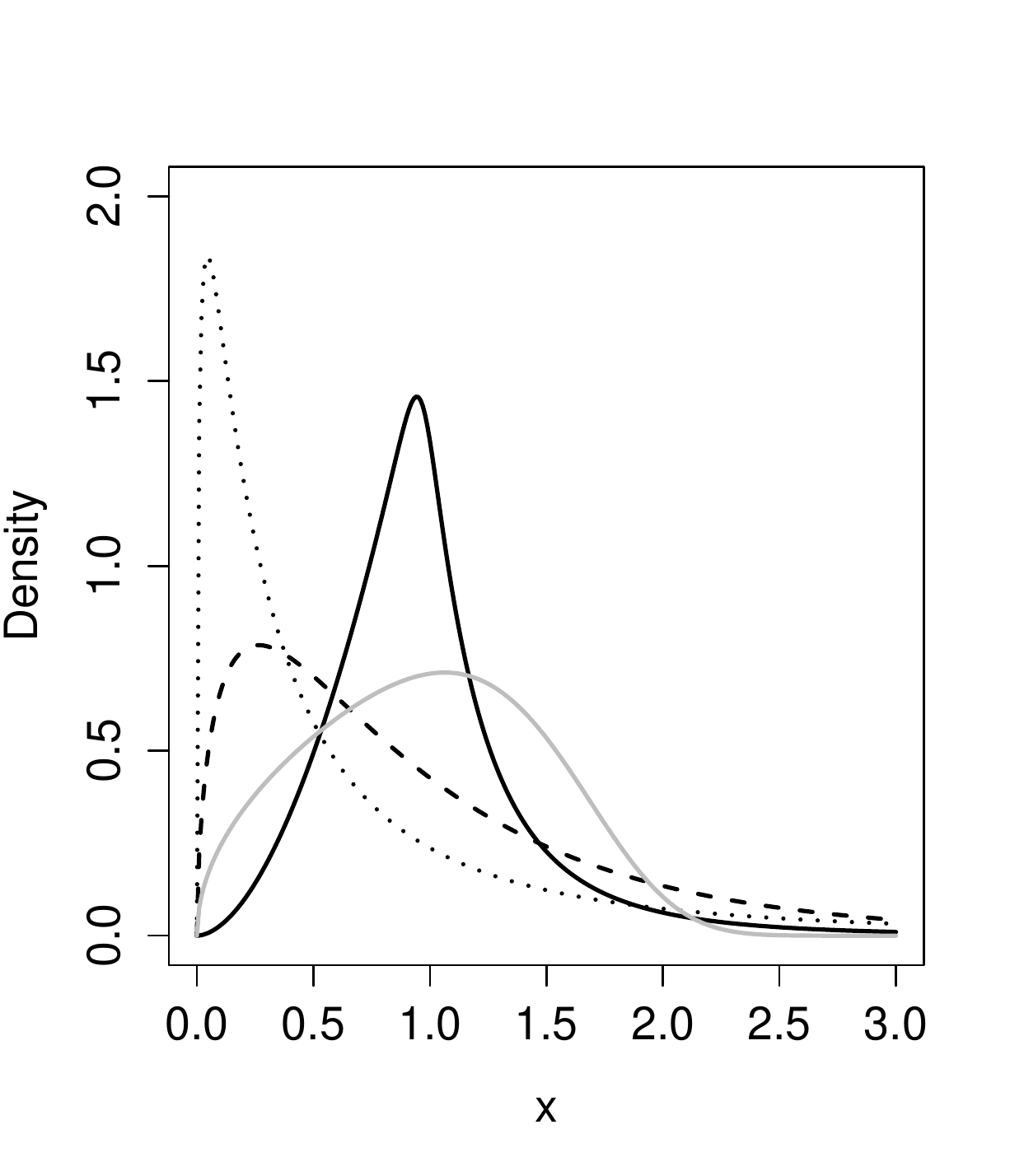}
\caption{The seven test densities.  Left Panel: Density 1 - solid; Density 2 - dashed; Density 3 - dotted. Right Panel: Density 4 - solid black; Density 5 - dashed; Density 6 - dotted; Density 7 - solid grey.} 
\label{fig:denssimul}
\end{figure}

In this section the practical performances of the Local Likelihood-based estimators are compared to those of their competitors. The same 7 test $\R^+$-supported densities as in \cite{Malec12} were considered. All belong to the family of generalised $F$-distributions \citep{Prentice75,McDonal84,Cox08}. This 4-parameter distribution is very flexible and includes as particular cases familiar distributions such as Exponential, Gamma and Generalised Gamma, Weibull, Log-Normal, and Singh-Maddala, among others. In particular, the generalised $F$-density can exhibit various behaviours close to 0 or in the tail. The seven densities are shown in Figure \ref{fig:denssimul}, for convenience. All have been parameterised for having expectation equal to 1. Note that Density 1 is the standard exponential density.

\ppn From each of these distributions, independent samples of size $n=100$ and $n=500$ have been generated, with $M=500$ Monte Carlo replications for each sample size. The densities were estimated from $x=0.001$ to $q_{0.999}$, the 99.9th quantile of each of them, on a grid of 1,000 values. Eight different estimators were considered:
\begin{itemize}\itemsep0em 
 \item[$\cdot$] \cite{Chen00}'s standard and modified Gamma kernel density estimators (in the tables below: `Gamma' and `Mod.Gam.', respectively);
 \item[$\cdot$] the `reflection' estimator \citep{Schuster85} (`Reflect.');
 \item[$\cdot$] the `Cut-and-Normalise' estimator \citep{Gasser79} (`CaN');
 \item[$\cdot$] the non-negative boundary-corrected kernel estimator of \cite{Jones96} (`Bound.');
 \item[$\cdot$] the `naive' log-transform estimator, i.e.\ the log-normal kernel density estimator (\ref{eqn:lognormkde}) (`Naive LT');
 \item[$\cdot$] the local log-quadratic transformation kernel density estimator ((\ref{eqn:ftildeTp}) with $p=2$) based on the log-transformation $T(x) = \log x$ (`LL-LT'); and
 \item[$\cdot$] the local log-quadratic transformation kernel density estimator ((\ref{eqn:ftildeTp}) with $p=2$) based on the `probex' transformation $T(x) = \Phi^{-1}(1-e^{-x})$ (`LL-PT').
\end{itemize}
For the two Gamma estimators, the bandwidth was chosen following the reference rule prescribed in \citet[Appendix B]{Hirukawa14}. For the reflection, Cut-and-Normalise and boundary estimators, the Gaussian kernel was used with a bandwidth selected by direct plug-in \citep{Sheather91}. These estimates were computed using the {\tt dbckden} function in the R package {\tt evmix}. For the log-normal kernel density estimator, the bandwidth was selected by direct plug-in, too, on the log-scale. Finally, the NN bandwidths $\alpha$ for the two local log-quadratic transformation kernel estimators were selected by the cross-validation procedure described in Section \ref{sec:bw}. The whole LLTKDE procedure (maximisation of (\ref{eqn:loclikpol}) and selection of $\alpha$) was implemented using the functions of the {\tt locfit} package in R. All estimates which require it, have been renormalised so as to integrate to 1 on $\R^+$.

\ppn No local log-linear ($p=1$) versions of (\ref{eqn:ftildeTp}) were considered, as the simulations in \cite{Geenens14} and \cite{Geenens16} clearly showed that the choice $p=2$ consistently produces better results. Likewise, the refined methods of \cite{Malec12} (Gamma estimators), \cite{Karunamuni05} (reflection) or \cite{Hirukawa14} (extra multiplicative bias correction) were not included in the simulation. Although those methods may slightly improve the performance compared to their basic versions, they are more complicated to implement. For instance, \cite{Malec12}'s `refined' Gamma estimator involves an extra tuning parameter whose selection requires pilot-estimating the density itself as well as its first two derivatives. 

\ppn For a given estimator $\hat{f}_X$, the Mean Integrated Absolute Relative Error (MIARE), i.e.\ $\E\left(\int_{0}^{\infty}\frac{|\hat{f}_{X}(x)-f_{X}(x)|}{f_{X}(x)}dx\right)$, was used as performance measure. This choice is justified here, as $f_X(x)$ may be unbounded at the boundary, and quickly tend to 0 in the tail. When integrating the estimation error over $\R^+$, an absolute error of $\epsilon$  when estimating $f_X$ where it is `very large' should not have the same impact as where it is `very small'. Hence, checking the relative estimation error seems more appropriate. The MIARE was approximated by
\begin{align*} \widehat{\text{MIARE}}\left(\hat{f}_{X}\right) & =  \frac{1}{M}\sum_{n=1}^{M} \sum_{i=1}^{1000}\frac{|\hat{f}_{X}\left(\frac{i \times q_{0.999}}{1000}\right)-f_{X}\left(\frac{i\times q_{0.999}}{1000}\right)|}{f_{X}\left(\frac{i\times q_{0.999}}{1000}\right)}, \end{align*}
where $M=500$ is the number of Monte Carlo replications. The results are reported in Table \ref{tab:full100} for $n=100$ and Table \ref{tab:full500} for $n=500$.

\ppn \begin{table}[h]
\centering
\begin{tabular}{r|rr|rrr|rrr}
  \hline
 \fbox{$n=100$} & \text{Gamma} & \text{Mod.\ Gam.}  & \text{Reflect.} & \text{CaN} & \text{Bound.}& \text{Naive LT} &  \text{LL-LT} &  \text{LL-PT} \\ 
  \hline
Dens.\ 1 &     0.495 &   0.466 &  0.679 &0.709 &0.693 &0.624 &0.484 & {\bf 0.269} \\
Dens.\ 2 &     1.809 &   1.389 &  1.826 &1.778 &1.674 &0.871 &{\bf 0.452} & 0.790 \\
Dens.\ 3 &    13.829 &  21.588 &  2.885 &4.125 &2.859 &0.581 &{\bf 0.378} & {\bf 0.534} \\
Dens.\ 4 &    11.184 &  16.482 &  4.539 &4.089 &2.738 &0.754 &{\bf 0.482} & {\bf 0.636} \\
Dens.\ 5 &     0.489 &   0.494 &  0.664 &0.723 &0.731 &0.581 &0.446 & {\bf 0.285} \\
Dens.\ 6 &     1.295 &   1.164 &  1.580 &1.585 &1.548 &0.786 &{\bf 0.434} & 0.621 \\
Dens.\ 7 &     0.508 &   0.727 &  {\bf 0.271} &{\bf 0.253} &{\bf 0.259} &0.396 &{\bf 0.288} & \bf{0.274} \\
   \hline
\end{tabular}
\caption{(approximated) MIARE, $n=100$. Bold values show the minimum MIARE for the corresponding density (non-significantly different values are highlighted as well). } 
\label{tab:full100}
\end{table}

\begin{table}[h]
\centering
\begin{tabular}{r|rr|rrr|rrr}
  \hline
 \fbox{$n=500$} & \text{Gamma} & \text{Mod.\ Gam.}  & \text{Reflect.} & \text{CaN} & \text{Bound.}& \text{Naive LT} &  \text{LL-LT} &  \text{LL-PT} \\ 
  \hline
Dens.\ 1 &     0.254  &  0.240 &  0.413 &0.459 &0.453 &0.343 &0.218 &   {\bf 0.117} \\
Dens.\ 2 &     1.508  &  1.147 &  1.950 &1.884 &1.812 &{\bf 0.447} &{\bf 0.464} &   0.780 \\
Dens.\ 3 &     5.105  &  7.506 &  1.350 &2.348 &1.317 &{\bf 0.338} &{\bf 0.229} &  {\bf 0.257} \\
Dens.\ 4 &     3.802  &  5.502 &  2.237 &2.282 &1.481 &0.455 &{\bf 0.301} &  {\bf 0.355} \\
Dens.\ 5 &     0.259  &  0.263 &  0.396 &0.462 &0.468 &0.313 &0.213 &   {\bf 0.153} \\
Dens.\ 6 &     0.866  &  0.786 &  1.370 &1.398 &1.389 &0.399 &{\bf 0.212} &   0.460 \\
Dens.\ 7 &     0.333  &  0.476 &  {\bf 0.153} &{\bf 0.140} &{\bf 0.140} & 0.230 &{\bf 0.142} &  {\bf  0.134} \\
   \hline
\end{tabular}
\caption{(approximated) MIARE, $n=500$. Bold values show the minimum MIARE for the corresponding density (non-significantly different values are highlighted as well). } 
\label{tab:full500}
\end{table}

\ppn For each density, the bold value is the minimum (approximated) MIARE (with non-significantly different values highlighted as well). All the bold values appear in the last two columns, i.e.\ the two LLTKD estimators (log- and probex-transforms), for all densities and sample sizes. This clearly evidences the superiority of these estimators compared to their competitors. The more conventional estimators (reflection, CaN and boundary-corrected) can only rival on Density 7, which has a very light right tail. The Gamma estimators are never matching the performance of the LLTKD estimators. Worse, for Densities 3 and 4, their estimated MIARE are huge. In fact, for those two densities, the Gamma estimators tend to produce estimates with too fat tails, similarly to what is seen in Figure \ref{fig:wage} below, increasing the relative error by a large amount. The two LLTKD estimators (log- and probex-transform) show very similar performance. As expected (see discussion in Section \ref{sec:logbey}), the estimator using the probex-transform does better for the Exponential density (Density 1), while the estimator based on the log-transform does better for densities similar to a log-normal, such as Density 6. All in all, they both perform very well. The `naive' estimator (\ref{eqn:lognormkde}) can not really rival the estimators based on Local Likelihood estimation, and this even when $f_X(x) \to 0$ as $x\to 0$ (Densities 3 and 4). Despite this, its performance increases with the sample size, and it often beats the conventional competitors as well.

\ppn It is also worth looking more closely at how the estimators behave in the right tail of the different densities. The above MIARE has thus been recomputed integrating over the values of $x$ over the 80th quantile of each test densities only, i.e., $\E\left(\int_{q_{0.80}}^{\infty}\frac{|\hat{f}_{X}(x)-f_{X}(x)|}{f_{X}(x)}dx\right)$ was approximated. The values $q_{0.80}$ are shown for each densities in Tables \ref{tab:tail100} and \ref{tab:tail500}, together with the approximated MIARE. The conclusion are very similar to what was said above. The LLTKD estimators prove superior in all cases, except for the very thin-tailed Density 7 for small samples (but they are not far behind).  The huge MIARE for the conventional estimators for Density 2 is implied by the occurrence of numerous `spurious bumps' in the estimated tails. In order to handle the peak that is unbounded at 0, the automatically selected bandwidth tends to be very small, hence producing highly undersmoothed estimates in the right tail. The same phenomenon occurs for Density 6, to a lesser extent.

\begin{table}[h]
\centering
\begin{tabular}{r|r|rr|rrr|rrr}
  \hline
 \fbox{$n=100$} & $q_{0.80}$ & \text{Gamma} & \text{Mod.\ Gam.}  & \text{Reflect.} & \text{CaN} & \text{Bound.}& \text{Naive LT} &  \text{LL-LT} &  \text{LL-PT} \\ 
  \hline
Dens.\ 1 &  2.306 &    0.673 &      0.633 &     0.933 &     0.970 &     0.955 &     0.862  &    0.646  &    {\bf 0.357}  \\
Dens.\ 2 &  0.560 &    1.831 &      1.402 &     1.847 &     1.800 &     1.696 &     0.883  &    {\bf 0.453}  &    0.798  \\
Dens.\ 3 &  1.393 &    2.456 &     5.442  &     1.070 &     0.892 &     0.895 &     0.736  &    {\bf 0.471}  &    {\bf 0.486}  \\
Dens.\ 4 &  1.489 &    0.892 &     1.235  &     1.341 &     1.279 &     1.282 &     0.885  &    {\bf 0.553}  &    {\bf 0.618} \\
Dens.\ 5 &  2.151 &    0.654 &     0.660  &     0.898 &     0.994 &     0.998 &     0.794  &    0.592  &    {\bf 0.366}  \\
Dens.\ 6 &  2.235 &    1.390 &     1.249  &     1.695 &     1.699 &     1.660 &     0.842  &    {\bf 0.463}  &    0.660  \\
Dens.\ 7 &  1.648 &    1.580 &     2.500  &     {\bf 0.562} &     {\bf 0.662} &     {\bf 0.673} &     1.058  &    0.710  &     0.706  \\
   \hline
\end{tabular}
\caption{(approximated) MIARE in the tail ($x > q_{0.80}$), $n=100$. Bold values show the minimum MIARE for the corresponding density (non-significantly different values are highlighted as well). } 
\label{tab:tail100}
\end{table}

\begin{table}[h]
\centering
\begin{tabular}{r|r|rr|rrr|rrr}
  \hline
 \fbox{$n=500$} & $q_{0.80}$ &  \text{Gamma} & \text{Mod.\ Gam.}  & \text{Reflect.} & \text{CaN} & \text{Bound.}& \text{Naive LT} &  \text{LL-LT} &  \text{LL-PT} \\ 
  \hline
Dens.\ 1 & 2.306 &      0.383 &     0.363 &     0.650 &     0.733 &     0.726 &     0.499 &     0.342 &     {\bf 0.146}   \\
Dens.\ 2 & 0.560 &      2.340 &     1.436 &    34.449 &    18.803 &    17.530 &     0.417 &     {\bf 0.201} &     0.273   \\
Dens.\ 3 & 1.393 &      0.532 &     0.904 &     0.855 &     0.595 &     0.595 &     0.416 &     {\bf 0.238} &     {\bf 0.270}   \\
Dens.\ 4 & 1.489 &      0.473 &     0.477 &     1.631 &     1.351 &     1.353 &     0.549 &     {\bf 0.271} &     {\bf 0.312}  \\
Dens.\ 5 & 2.151 &      0.372 &     0.367 &     0.591 &     0.726 &     0.733 &     0.440 &     0.336 &     {\bf 0.168}   \\
Dens.\ 6 & 2.235 &      1.060 &     0.901 &     3.959 &     4.172 &     4.122 &     0.437 &     {\bf 0.205} &     0.328  \\
Dens.\ 7 & 1.648 &      1.019 &     1.578 &     {\bf 0.279} &     {\bf 0.319} &     {\bf 0.324} &    0.592 &     {\bf 0.313} &    {\bf 0.325}   \\
   \hline
\end{tabular}
\caption{(approximated) MIARE in the tail ($x > q_{0.80}$), $n=500$. Bold values show the minimum MIARE for the corresponding density (non-significantly different values are highlighted as well). } 
\label{tab:tail500}
\end{table}

\section{Real data analyses} \label{sec:realdata}

In this section, three real data sets are analysed. The $\R^+$-supported densities of the distributions that have generated them are estimated using the local log-quadratic transformation kernel estimator (with probex transformation and NN-bandwidth $\alpha$) described in the previous sections. For comparison, \cite{Chen00}'s two Gamma kernel estimators as well as \cite{Jones96}'s boundary-corrected kernel estimator are shown, together with an histogram and the data themselves (ticks at the bottom of the plots). The smoothing parameters for the Gamma estimators were chosen by the Gamma reference rule \citep{Hirukawa14} and that for the boundary-kernel estimator by direct plug-in \citep{Sheather91}. Of course, the NN-bandwidth $\alpha$ was selected as per Section \ref{sec:bw}. Importantly, the three data sets were first rescaled so as to have their average equal to 1, this to allow the transformation to play the role that it is intended to fill.\footnote{The straight application of (\ref{eqn:probex}) to values in their 100's or 1,000's, as those observed in the three datasets, would essentially send all of them to $+\infty$, which is not the purpose.} Of course, the density estimates are shown below in their original scale.

\begin{figure}[h]
\centering
\includegraphics[width=0.9\textwidth]{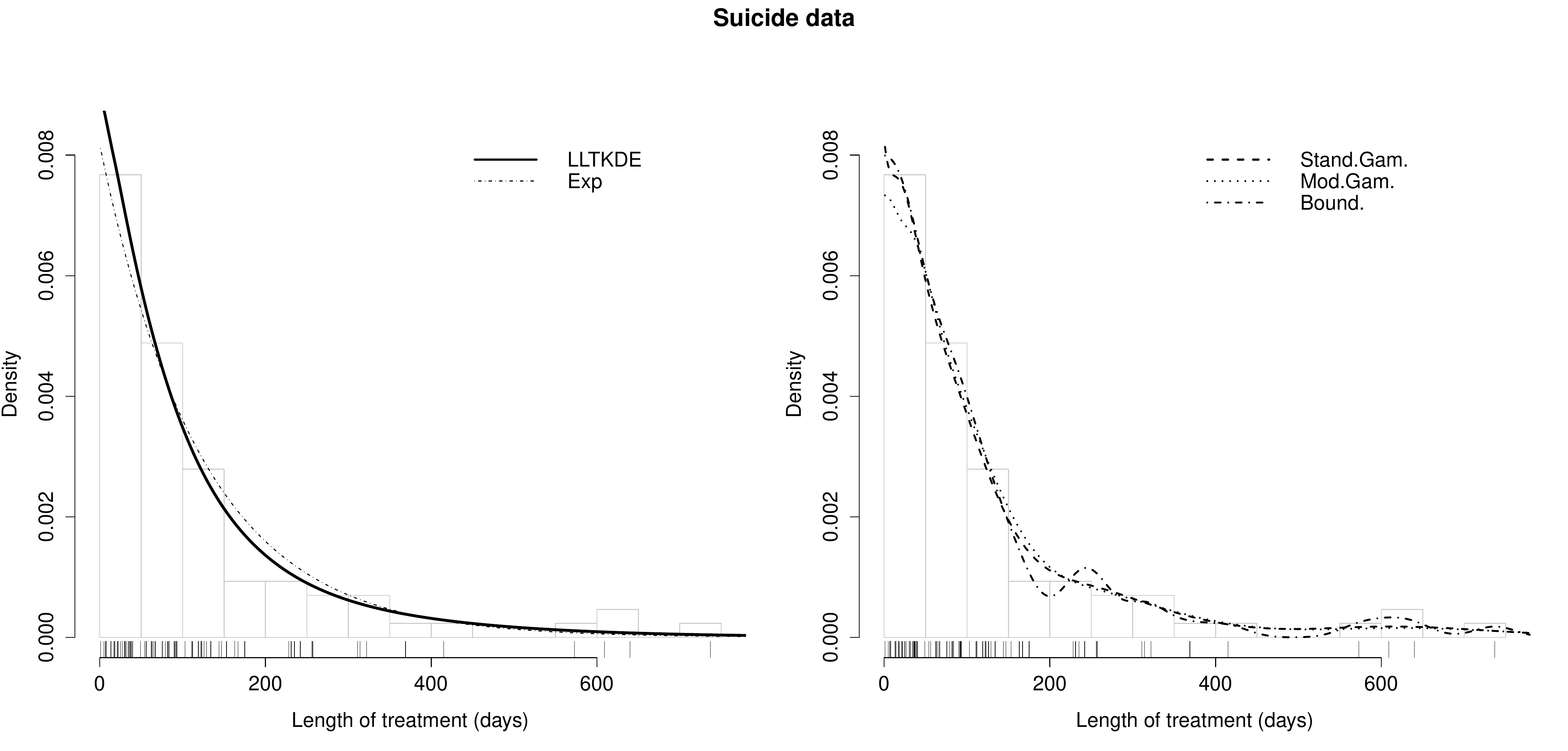}
\caption{`Suicide' data set: local log-quadratic probex transformation estimate and fitted Exponential density (left panel), Gamma estimators and boundary-corrected kernel estimator (right panel).}
\label{fig:suicide}
\end{figure}

\ppn The first data set is the `suicide' data set, which gives the lengths (in days) of $n=86$ spells of psychiatric treatment undergone by patients used as controls in a study of suicide risks. It is originally reported by \cite{Copas80}, and is also studied among others in \cite{Silverman86} and \cite{Chen00} in relation to boundary correction problems. The estimation of suicide risk as a function of time under psychiatric treatment has attracted some attention in the psychiatric literature in the past, as it is important to predict the risk such that suitable treatment can be made for the patient. Figure \ref{fig:suicide} show the different density estimates. On the left panel, the local log-quadratic transformation estimate, with $\alpha = 0.97$ automatically picked by LSCV (\ref{eqn:LSCV}), is shown. It is actually very close to an Exponential density (the Maximum Likelihood Exponential density is shown, too, for convenience), and shares with it the same smoothness and visual appeal. This is actually a power of Local Likelihood density estimation methods: they are purely nonparametric but they can also act `semi-parametrically' in some sense when the bandwidth is kept large \citep{Eguchi98,Park06}. Here, the LSCV smoothing parameter $\alpha \simeq 1$ is indeed `large'. Loosely speaking, the procedure has `felt' that the underlying density was close to Exponential, so picked $\alpha$ large so as to guide the resulting estimate toward that density shape. This is, of course, a consequence of working with the `probex' transformation: in the transformed domain, $f_Y$ should be close to Normal, and this is essentially what the local log-quadratic estimate would look like for large $\alpha$. 

\ppn The smoothing parameter for the Gamma estimators was selected according to a Gamma reference distribution, so this should be close to optimal here as the Exponential is a particular case of the Gamma. Yet, the behaviour of the two estimates at the boundary is not satisfactory. The standard Gamma shows an inelegant kink, whereas the modified Gamma seems to underestimate $f_X$ there, compared to the other estimates and the histogram. This is typical of the modified Gamma estimator, as discussed in \cite{Malec12}. The boundary-corrected kernel estimate is doing reasonably at the boundary, but exhibits numerous `spurious bumps' in the right tail for $x>200$.

\begin{figure}[h]
\centering
\includegraphics[width=0.9\textwidth]{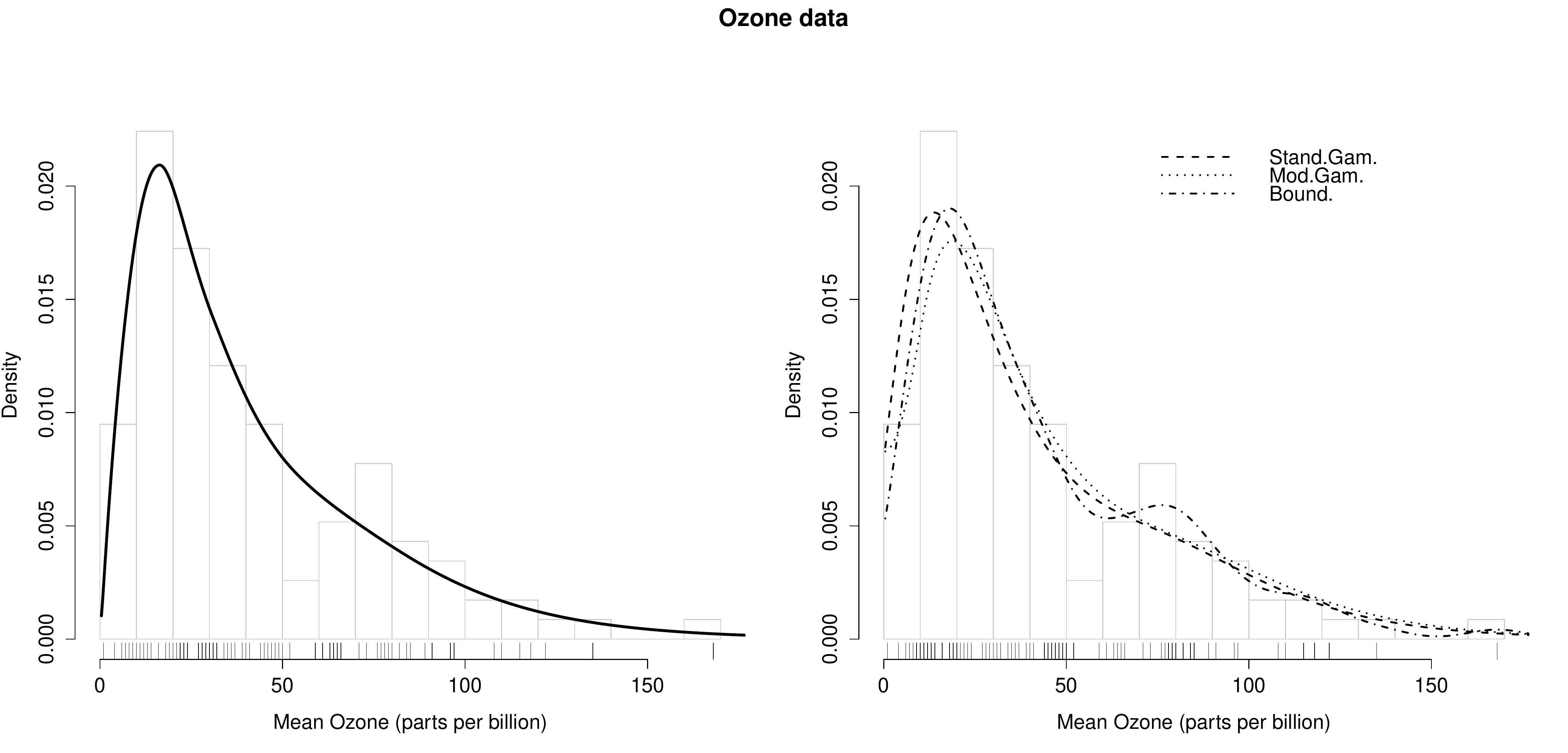}
\caption{`Ozone' data set: local log-quadratic probex transformation estimate (left panel), Gamma estimators and boundary-corrected kernel estimator (right panel).}
\label{fig:ozone}
\end{figure}

\ppn The second data set contains the mean ozone concentration (in parts per billion) in the air from 13:00 to 15:00 hours at Roosevelt Island, New York City, from May 1, 1973 to September 30, 1973. it was obtained from the New York State Department of Conservation.\footnote{This sample is actually part of the {\tt airquality} data set available from R.} After the missing values are removed, there remain $n=116$ observations. Importantly, given that the ozone concentration was measured day after day, these data actually form a time series, which is not a problem as per Subsection \ref{subsec:dependent}. The value of $\alpha$ picked by (\ref{eqn:LSCV}) is here $\alpha=0.91$, which gives the LLTKD estimate in the left panel of Figure \ref{fig:ozone}. Again, the estimate is rightly smooth all over $\R^+$. It picks the high peak at around $x=20$ better than its competitors (right panel), while appropriately plunging to 0 as $x \to 0$, unlike the others. It also estimates the fat right tail without problem.

\begin{figure}[H]
\centering
\includegraphics[width=0.9\textwidth]{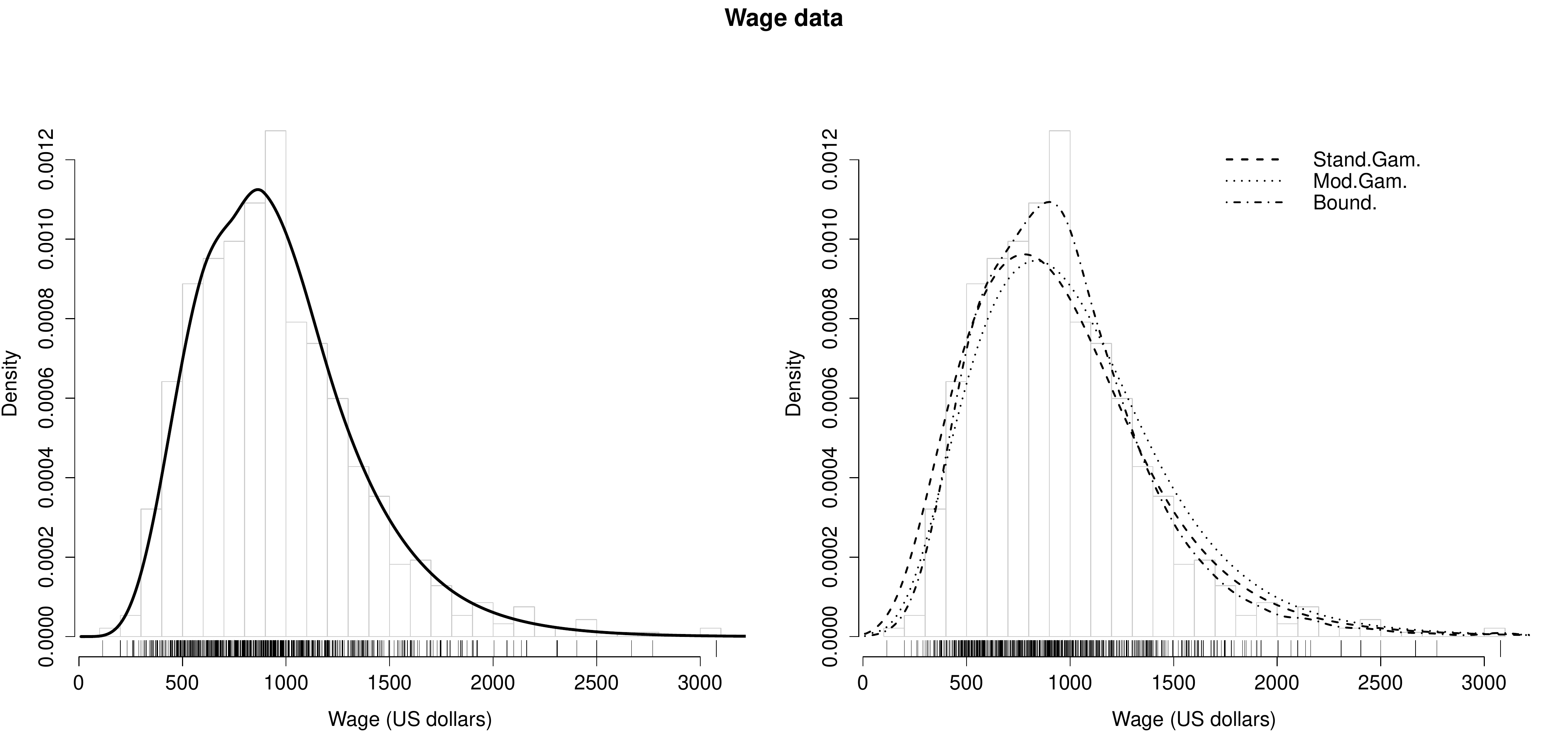}
\caption{`Wage2' data set: local log-quadratic probex transformation estimate (left panel), Gamma estimators and boundary-corrected kernel estimator (right panel).}
\label{fig:wage}
\end{figure}

\ppn Finally, Figure \ref{fig:wage} shows density estimates for the `wage2' dataset, available from \cite{Wooldridge13} and already examined in the context of kernel density estimation in \cite{Hirukawa14}. It gives monthly earnings of $n=935$ US males (in US dollars). The value $\alpha= 0.75$ was automatically selected by LSCV. This example does not pose any particular problem but is presented here to show that the local log-quadratic estimator performs as well for densities tending to 0 very smoothly as $x \to 0$, as for the previously considered cases. In all situations, the obtained LLTKD estimates are smooth and visually appealing, but do not oversmooth important features of the underlying densities. This is a pleasant feature of this type of estimators which was already praised in \cite{Geenens14} and \cite{Geenens16}. In fact, the reduction of the bias order allows the actual amount of smoothing applied to the data to be increased, without seeing the bias growing too much. The final estimates are thus generally smoother than their `conventional' kernel counterparts. The cross-validation criterion (\ref{eqn:LSCV}) consistently picks a value of $\alpha$ which achieves an optimal bias-variance trade-off in practice.

\section{Concluding remarks} \label{sec:ccl}

In this paper, another look at the transformation idea for estimating $\R^+$-supported densities was given, in spite of the bad reviews it had got in the previous literature. A key observation was that, the best estimate of the density of the transformed random variable $Y = T(X)$ does not necessarily produce the best estimate of the density of $X$ after back-transformation through $f_X(x) = f_Y(T(x)) \times T'(x)$. Specifically, here, it was seen that particular care should be given to estimating the left tail of $f_Y$ more smoothly and accurately than other parts. Natural candidates for achieving this were the local-likelihood density estimators, known to have superior tail behaviour. Combining the transformation and such local-likelihood estimation has lead to very good density estimates, especially when used in conjunction with a Nearest-neighbour type of bandwidth. A practical way for selecting that bandwidth was devised, and was seen to perform very well. Beyond attractive theoretical properties, the so-produced estimators have shown practical appeal, too. In particular, the estimates are typically smooth all over $\R^+$ and visually pleasant, but without oversmoothing. This confirms the observations made in \cite{Geenens14} and \cite{Geenens16} for similar estimators in different contexts.

\ppn The proposed methodology is based on 4 main ingredients: $i$) transformation, $ii$) local-likelihood estimation, $iii$) Nearest-Neighbour bandwidth, $iv$) cross-validation. Taken separately, these four procedures have been criticised in the earlier literature: $i$) the transformation idea was seen not to perform as expected (see Section \ref{subsec:naive}); $ii$) the local-likelihood estimators were seen not to match the raw kernel density estimator in terms of MISE in general \citep{Hall02}; $iii$) Nearest-Neighbour bandwidths are known to produce rough, fat-tailed and not integrable estimates when used with conventional kernel density estimation; and $iv$) the least-squares cross-validation criterion was blamed for being very unstable and typically leading to undersmoothing. Interestingly, though, combining $i$)-$ii$)-$iii$)-$iv$) together seems to produce excellent results, as it is evidenced in this paper.

\ppn An important feature of the suggested estimators is their aptitude for accurately estimating the right tail of the density of interest, as was demonstrated by the simulation study (Section \ref{sec:simul}). In the case where the right-tail behaviour of the density is the real focus of the analysis, for instance when a high quantile should be estimated, they could really play an important role in future developments. For instance, it is expected that the LLTKDE could produce better estimates of Value-at-Risk, or Expected Shortfall, of a loss distribution. This problem will be studied in depth in a forthcoming paper.

\section*{Acknowledgements} 
This research was supported by a Faculty Research Grant from the Faculty of Science, UNSW Australia, Sydney (Australia).

\end{document}